\documentclass{pasj01} 

\Received{$\langle$reception date$\rangle$}
\Accepted{$\langle$acception date$\rangle$}
\Published{$\langle$publication date$\rangle$}
\usepackage{graphicx}
\usepackage{tabularx}
\usepackage{natbib}
   \newcolumntype{C}{>{\centering\arraybackslash}X}
   \newcolumntype{L}{>{\raggedright\arraybackslash}X}
   \newcolumntype{R}{>{\raggedleft\arraybackslash}X}
   \renewcommand{\arraystretch}{1.2}
\begin{document}

\title{Spatially resolved study of the SS 433/W50 west region with Chandra: X-ray structure and spectral variation of non-thermal emission}
\author{Kazuho \textsc{Kayama},\altaffilmark{1,}$^{*}$ 
Takaaki  \textsc{Tanaka},\altaffilmark{2, 1}
Hiroyuki \textsc{Uchida},\altaffilmark{1}
Takeshi Go \textsc{Tsuru},\altaffilmark{1} 
Takahiro \textsc{Sudoh},\altaffilmark{3, 4}
Yoshiyuki \textsc{Inoue},\altaffilmark{5, 6, 7}
Dmitry \textsc{Khangulyan},\altaffilmark{8}
Naomi \textsc{Tsuji},\altaffilmark{9, 6, 8}
Hiroaki \textsc{Yamamoto}\altaffilmark{10}
}%
\altaffiltext{1}{Department of Physics, Kyoto University, Kitashirakawa Oiwake-cho, Sakyo, Kyoto 606-8502, Japan}
\altaffiltext{2}{Department of Physics, Konan University, 8-9-1 Okamoto, Higashinada, Kobe, Hyogo 658-8501, Japan}
\altaffiltext{3}{Department of Astronomy, University of Tokyo, Hongo, Tokyo 113-0033, Japan}
\altaffiltext{4}{Center for Cosmology and AstroParticle Physics, Physics Research Building, Ohio State University, 191 West Woodruff Avenue, Columbus, OH 43210, USA}
\altaffiltext{5}{Department of Earth and Space Science, Graduate School of Science, Osaka University, Toyonaka, Osaka 560-0043, Japan}
\altaffiltext{6}{Interdisciplinary Theoretical \& Mathematical Science Program (iTHEMS), RIKEN, 2-1 Hirosawa, Saitama 351-0198, Japan}
\altaffiltext{7}{Kavli Institute for the Physics and Mathematics of the Universe (WPI), UTIAS, The University of Tokyo, Kashiwa, Chiba 277-8583, Japan}
\altaffiltext{8}{Department of Physics, Rikkyo University, Nishi-Ikebukuro 3-34-1, Toshima-ku, Tokyo 171-8501, Japan}
\altaffiltext{9}{Faculty of Science, Kanagawa University, 2946 Tsuchiya, Hiratsuka-shi, Kanagawa 259-1293, Japan}
\altaffiltext{10}{Department of Astrophysics, Nagoya University, Chikusa-ku, Nagoya, Aichi 464-8602, Japan}

\email{kayama.kazuho.57r@kyoto-u.jp}

\KeyWords{X-rays: binaries${}_1$ --- radiation mechanisms: non-thermal${}_2$ --- ISM: jets and outflows${}_3$}

\maketitle

\begin{abstract}
The X-ray binary SS~433, embedded in the W50 nebula (or supernova remnant W50), shows bipolar jets that are ejected with mildly relativistic velocities, and extend toward the east and west out to scales of tens of parsecs.
Previous X-ray observations revealed twin lobes along the jet precession axis that contain compact bright knots dominated by synchrotron radiation, which provide evidence of electron acceleration in this system. 
Particle acceleration in this system is substantiated by the recently detected gamma rays with energies up to at least 25 TeV.
To further elucidate the origin of the knots and particle acceleration sites in SS~433/W50, we report here on detailed, spatially resolved X-ray spectroscopy of its western lobe with Chandra.
We detect synchrotron emission along the jet precession axis, as well as optically thin thermal emission that is more spatially extended.
Between the two previously known knots, w1 and w2, we discover another synchrotron knot, which we call w1.5.
We find no significant synchrotron emission between SS~433 and the innermost X-ray knot (w1), suggesting that electrons only begin to be accelerated at w1.
The X-ray spectra become gradually steeper from w1 to w2, and then rapidly so immediately outside of w2.
Comparing with a model taking into account electron transport and cooling along the jet, this result indicates that the magnetic field in w2 is substantially enhanced, which also explains its brightness.
We discuss possible origins of the enhanced magnetic field of w2 as well as scenarios to explain the other two knots.
\end{abstract}

\section{Introduction}
SS~433 is a well-known Galactic microquasar, which is a binary system hosting a black hole (or a neutron star) and an A-type supergiant with an orbital period of 13.1~days \citep{ref_Hillwig_2004}.
A pair of jets ejected toward the east and west from SS~433 is measured to have a mildly relativistic velocity, 0.26$c$ at the jet base \citep{ref_margon_1989}, where $c$ is the speed of light.
The jets propagate into the surrounding nebula W50, whose origin is under debate \citep[e.g., ][]{ref_goodall_2011,ref_ohmura_2021}.
X-ray observations of SS~433/W50 revealed extended east-west lobes \citep[e.g.,][]{ref_watson_1983,ref_brinkmann_1996} and compact bright knots \citep{ref_safi-harb_1997} along the precession axis of both jets.
While the origins of these X-ray structures are still unclear, the spatial distributions suggest that they reflect the past activity of the jets and thus SS~433 itself.

X-ray spectra of the lobes are well fitted with power laws \citep{ref_yamauchi_1994,ref_safi-harb_1999}.
The emission is attributed to synchrotron radiation from relativistic electrons, suggesting that particle acceleration is at work in the SS 433/W50 system. 
Particle acceleration in this system is further evidenced by recent gamma-ray observations. 
\cite{ref_abeysekara_2018} reported a detection of very-high-energy (VHE) gamma-ray emissions with the High Altitude Water Cherenkov Observatory (HAWC) from the western and eastern lobes. 
The VHE emission extends up to at least 25~TeV, indicating that particles, either electrons or protons, are accelerated to $\gtrsim 100~\mathrm{TeV}$. 
Several authors reported analysis results of GeV gamma-ray observations with Fermi-LAT \citep[][]{ref_bordas_2015,ref_rasul_2019,ref_xing_2019,ref_sun_2019,ref_fang_2020,ref_li_2020}.
However, the conclusions of these analyses, including those on the emitting regions and spectra, are inconsistent with each other.

Several theoretical models were proposed to account for the radio-to-gamma-ray emissions found along the jet precession axis of SS~433/W50. 
\cite{ref_sudoh_2020} constructed a leptonic model, in which they assume electrons accelerated near the innermost knots; in this scenario, the magnetic fields should be enhanced at the knots.
\cite{ref_kimura_2020} tested both leptonic and hadronic scenarios, and concluded that the hadronic models are also able to reproduce the observed multi-wavelength spectra.
The origin of X-ray lobes, including the knots, will provide us with a clue to understanding their radiation mechanism in relation to the jet activity.
Previous X-ray studies focused mainly on spectra extracted from regions with relatively large angular scales \citep[e.g.,][]{ref_yamauchi_1994,ref_safi-harb_1999,ref_namiki_2000,ref_brinkmann_2007}.
Spatially resolved X-ray spectroscopy is essential to disentangle possible scenarios about the radiation mechanisms and particle acceleration in SS~433/W50. 

In this work, we report on a detailed spatially resolved spectroscopy of the W50 western lobe (knots) with the Chandra X-ray Observatory.
The high angular resolution of Chandra allows us to reveal a spectral variation along the jet precession axis on a few arcmin scales.
We compare our result with a theoretical calculation to discuss the nature of the X-ray emission from the knots and the origin of W50.
Throughout this paper, we assume 5.5~kpc as the distance to SS~433 \citep{ref_hjellming_1981,ref_blundell_2004,ref_lockman_2007}, though another possibility ($\sim3.8$~kpc) is claimed based on a parallax measurement with Gaia \citep{ref_arnason_2021}.
Through this paper, statistical uncertainties are quoted at $\rm 1\sigma$ confidence intervals.

\section{Observation and Analysis}
\subsection{Data Reduction}
We analyzed Chandra ACIS data of the western lobe of W50 taken in 2003 August (ObsID=3843).
The observation was carried out with I0, I1, I2, and I3 chip from the ACIS-I and S2 and S3 chip from the ACIS-S.
The data were reduced using the software package CIAO 4.12 and version 4.9.1 of the calibration database.
The effective exposure time of the observation is 66.7~ks after filtering out time intervals affected by background flares. 

\subsection{Imaging Analysis}
\label{sec_imaging_ana}

\begin{figure}
 \begin{center}
\includegraphics[width=75mm]{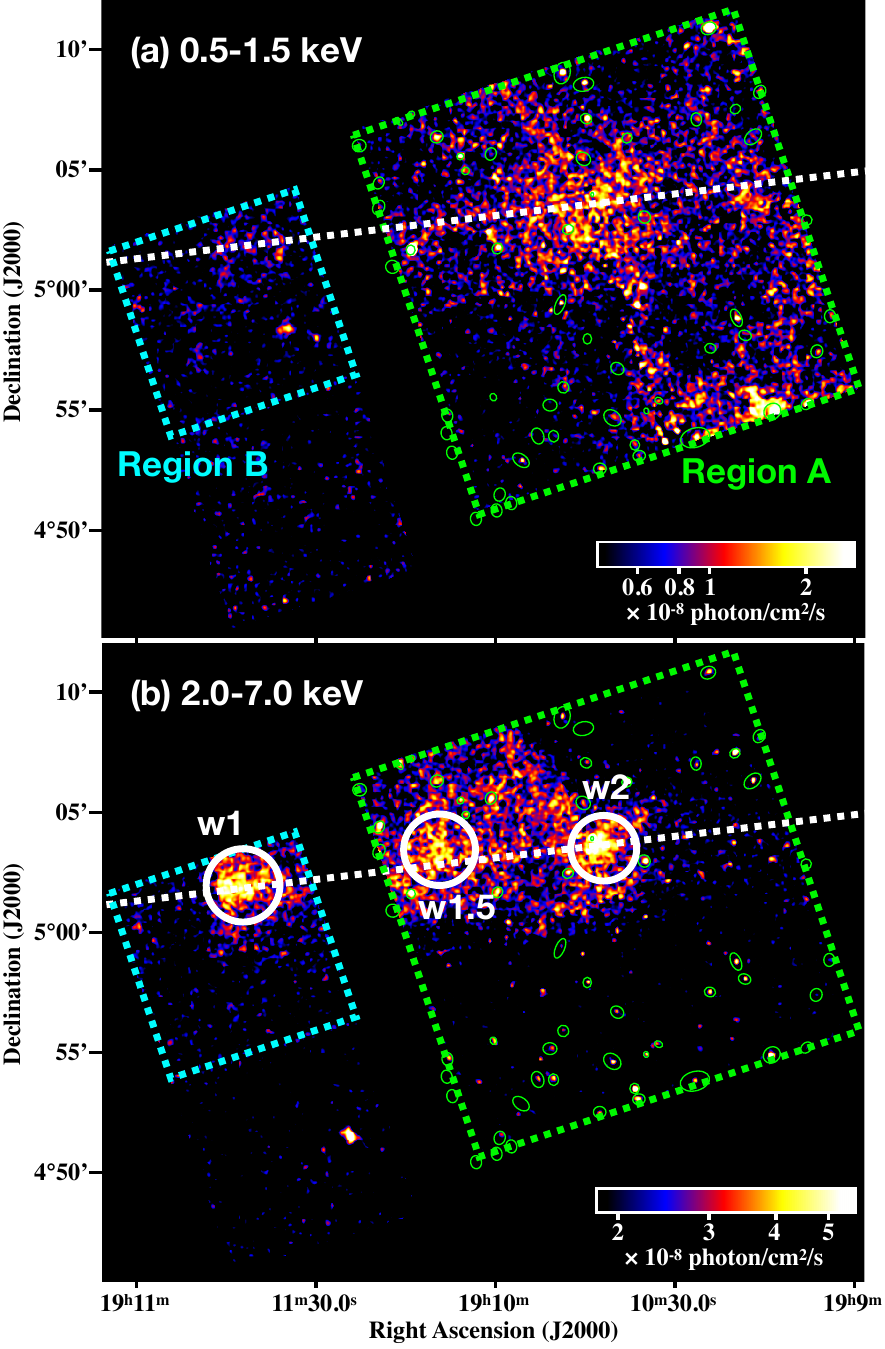} 
\end{center}
\caption{Chandra ACIS image of the western lobe of W50 in the energy bands of (a) 0.5--1.5 keV and (b) 2.0--7.0 keV.
The white dashed line corresponds to the jet precession axis determined by \cite{ref_eikenberry_2001}. 
The locations of the knots w1, w2, and w1.5 are marked with the white circles.
The green and cyan dashed rectangles indicate Regions~A and B, covered by the FI and BI chips, respectively.
The green circles indicate the locations of point sources detected with \texttt{wavdetect}.}
\label{fig_img_b}
\end{figure} 

Figure~\ref{fig_img_b} shows exposure-corrected images of the western lobe of W50. 
The particle background is estimated and subtracted from the images by referring to the prescription using ACIS-stowed observations by \cite{ref_hickox_2006}. 
The soft X-ray distribution seems more extended whereas hard X-rays are visible only around the jet precession axis of SS~433.
In addition to the two knots, w1 and w2 as named by \cite{ref_safi-harb_1997}, we discovered another bright spot in between. 
We name the new knot ``w1.5'' as displayed in figure~\ref{fig_img_b}. 
We note that the definitions of the knot w1 and w2 are different from \cite{ref_safi-harb_1997} and have been redefined based on our Chandra images.
A number of point sources are detected in the field of view (FOV) with the \texttt{wavdetect} software. 
We excluded the point sources in the spectral analysis of the extended emission described below. 

\begin{figure*}
\begin{center}
\includegraphics[width=170mm]{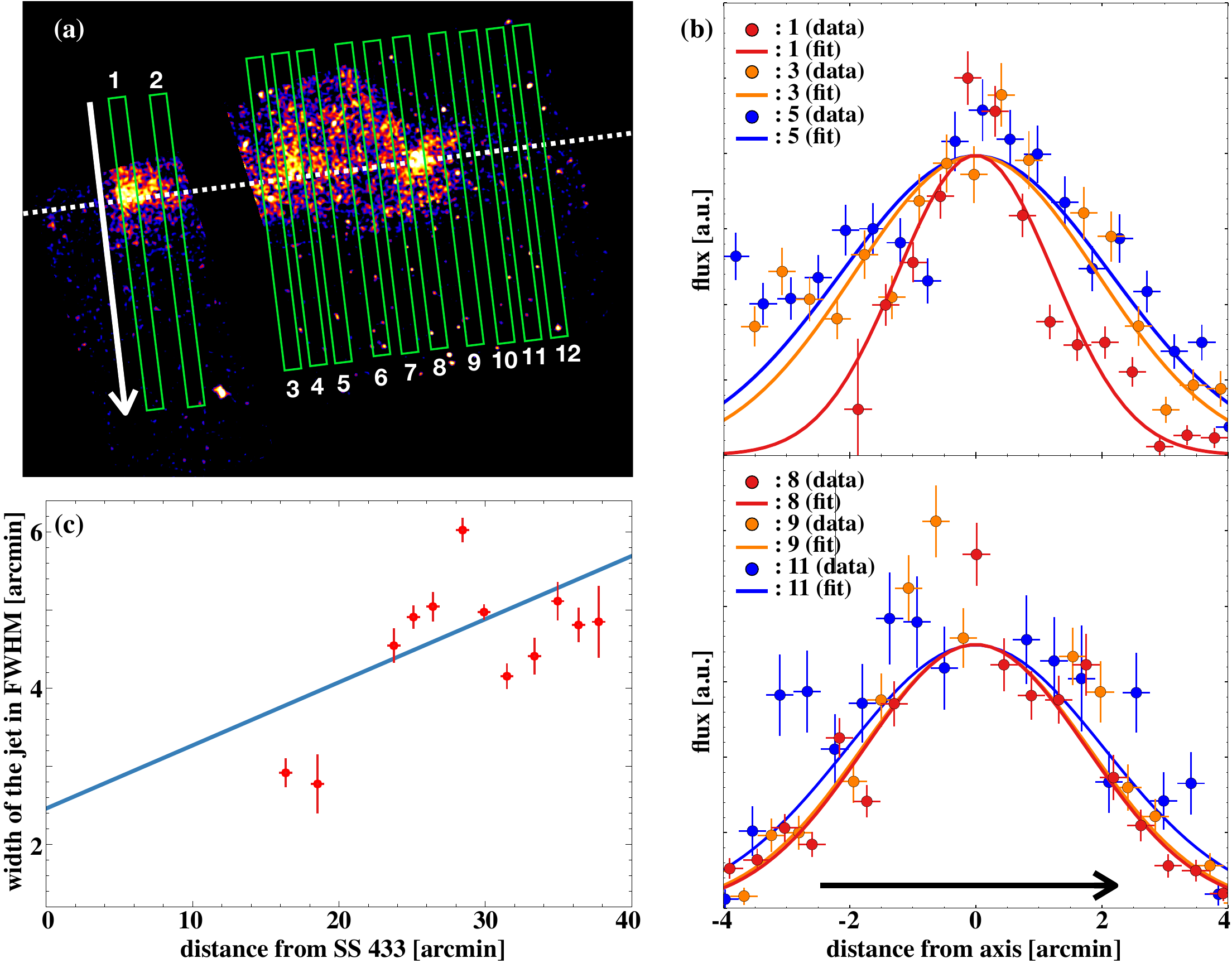} 
\end{center}
\caption{(a) Chandra ACIS image of 2.0--7.0 keV X-rays. The green rectangles are regions used for the jet opening angle measurement. 
The white dashed line indicates the jet precession axis. 
(b) The profiles in some representative regions in figure 2a. The numbers in the panels denote regions in figure 2a.
The arrows in figures 2a and 2b indicate the direction of the projection. 
(c) Widths of the jet in FWHM as a function of angular distance from SS~433. The blue line shows the result of linear regression of the data points.
}\label{fig_openangle}
\end{figure*}

We quantify the geometry of the western jet based on the hard band image in figure~\ref{fig_img_b}b.
Half-opening widths of the jet are estimated from the profiles in each rectangular region in figure~\ref{fig_openangle}a.
We fit the profiles with a phenomenological model composed of a Gaussian and a constant, corresponding to the emission from the jet and sky background components, respectively.
Both data and fitting results, from which the background contribution is subtracted, are shown in figure~\ref{fig_openangle}b.
Note that a part of the emission may be outside the FOV, but the majority, including the intensity peaks, are within the FOV.
Figure~\ref{fig_openangle}c presents the obtained widths in full width at half maximum (FWHM) plotted against angular distance from SS~433.
The width increases with the distance, implying that the jet is conical rather than cylindrical.
This agrees with the suggestions by previous observations and with assumptions often made in the literature \citep[e.g.,][]{ref_goodall_2011,ref_sudoh_2020}.
Linear regression of the data points in figure~\ref{fig_openangle}c gives a jet half-opening angle of $\timeform{2.6D}\pm\timeform{0.3D}$. 

\subsection{Spectral Analysis}
\subsubsection{Global Spectra}
\label{sec_analysis_1}
We first analyze spectra spatially integrated over the whole FOV of the ACIS. 
The background cannot be extracted from the present data since X-ray emission presumably from W50 is spread over the FOV (figure~\ref{fig_img_b}).
We instead model the background as detailed below.
Extracted spectra are composed of particle and sky backgrounds in addition to the emission from W50.
Since the particle background is different between the Front-Illuminated (FI) and Back-Illuminated (BI) chips, we separately analyze ACIS-I and S3 spectra to determine their background levels. 
The sky background is modeled by referring to studies of the Galactic backgrounds and cosmic X-ray background (CXB) in the literature.

\begin{figure}
 \begin{center}
\includegraphics[width=80mm]{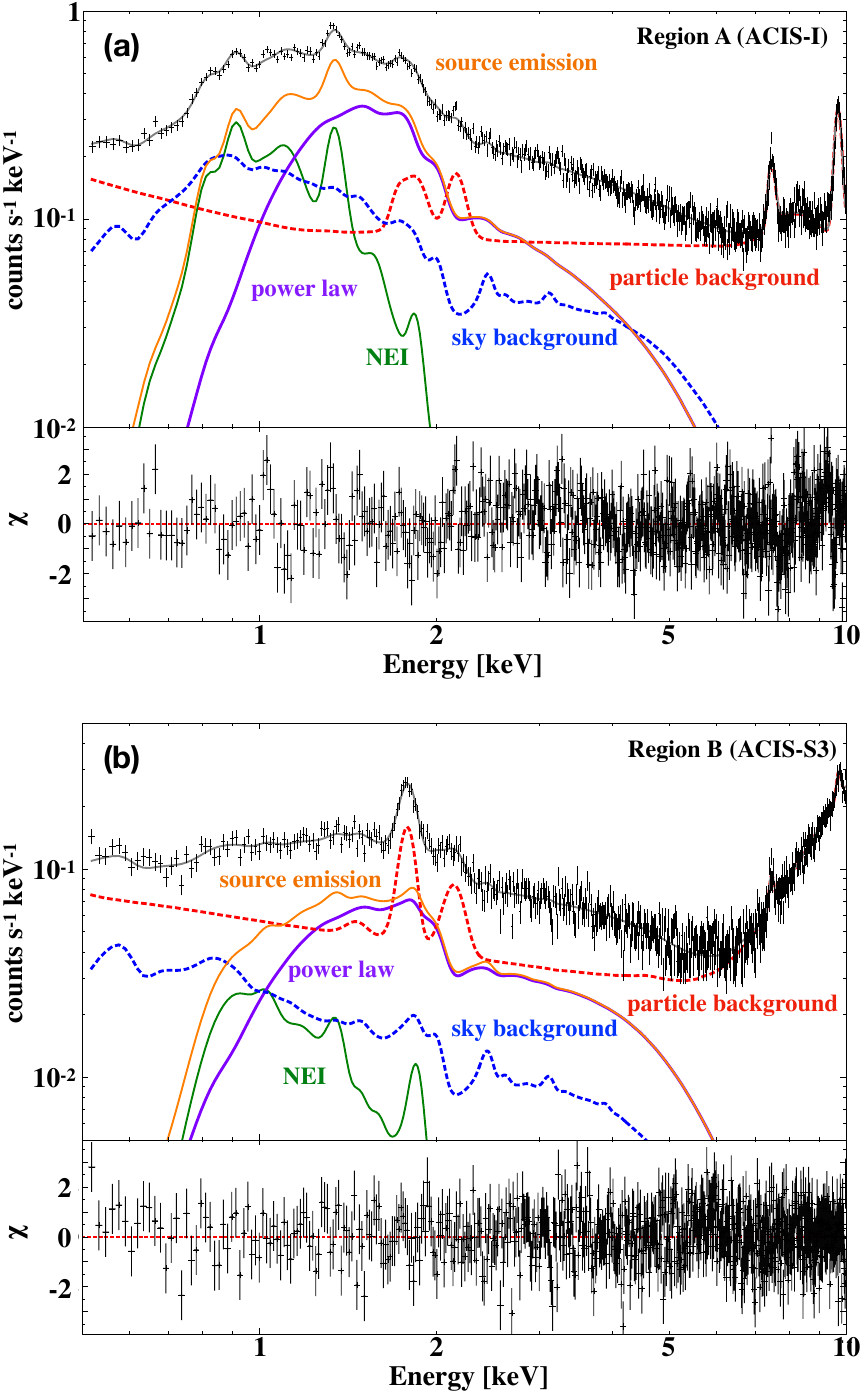} 
\end{center}
\caption{Spectra extracted from Region~A (a) and Region~B (b).
The sky- and particle-background models are indicated by the blue and red lines, respectively. 
The thermal (NEI) and non-thermal (power law) models are shown by the green and purple lines, respectively.
The orange lines indicate the total source emission models.
The bottom panels present residuals from the best-fit models.
}\label{fig_entire_spectrum}
\end{figure}

Figure~\ref{fig_entire_spectrum} shows spectra obtained with the FI and BI chips from Regions~A and B, respectively.
Following spectral fits were performed with XSPEC version 12.10.1f. 
The particle background models are based on the works by \cite{ref_bartalucci_2014} for FI and by \cite{ref_sharda_2020} for BI: 
an exponential function with a power law for Region~A and a three-segment broken power-law with a broad Gaussian for Region~B.
We allow to vary the normalizations of these components.
We model with five Gaussians the fluorescence lines of Si~K, Au~M, Au~L, and Ni~K emitted from the materials of the instrument.
Note that their mean energies are higher than the literature values because of an excessive charge transfer inefficiency correction for the pulse heights of the events for the fluorescence lines \citep{ref_grant_2005}.
We thus set the mean energies, line widths, and normalizations of the Gaussians also as free parameters.

The sky background consists of the Galactic Ridge X-ray Emission (GRXE), CXB, and Foreground Emission (FE) \citep[e.g.,][]{ref_uchiyama_2013}.
Since W50 is located near the Galactic plane in the inner Galaxy region, the contribution of the GRXE is not negligible \citep{ref_koyama_1996,ref_ebisawa_2008}.
The GRXE can be explained as emissions from low- and high-temperature plasmas \citep[LP and HP, respectively;][]{ref_uchiyama_2013} that are in collisional ionization equilibrium \citep[APEC;][]{ref_smith_2001}.
Based on the distribution of the GRXE given by \cite{ref_uchiyama_2013}, we set the normalizations of HP and LP at the coordinates of the FOV: $(l, b)=($\timeform{39.43D}, \timeform{-1.82D}$)$. 
The CXB is assumed to be a power law with a photon index of $\Gamma = 1.4$, and its normalization is fixed to $\rm 6.38\times10^{-8}~erg~cm^{-2}~s^{-1}~sr^{-1}$ in the energy band of 2$-$10~keV  \citep{ref_kushino_2002}.
Since the GRXE is a subdominant component in the low-energy band, the absorption column density ($N_{\rm H}$) cannot be constrained.
Therefore, we fix the $N_{\rm H}$ to the value of the Galactic ridge described by \cite{ref_uchiyama_2013}.
Since the effective distance to the Galactic plane is roughly equal to half the depth of the Galactic disk along the line of sight, the CXB photons are thought to propagate, inside the Galaxy, twice the distance of GRXE.
We thus set the $N_{\rm H}$ of CXB to twice that of the GRXE as in several previous studies of Galactic sources \citep[e.g.,][]{ref_nakashima_2013, ref_suzuki_2020}.
The spectra also have a predominant component called FE in the energy band below 1.2 keV \citep[e.g.,][]{ref_ryu_2009,ref_uchiyama_2013}.
The origin of this emission is unclear but is considered either a local Galactic plasma \citep{ref_ebisawa_2008} or unresolved faint dM stars \citep{ref_masui_2009}.
The FE component is modeled with two APEC ($\rm FE_{low}$ and $\rm FE_{high}$) components to which we apply parameters given by \cite{ref_uchiyama_2013}.
The ratio of these normalizations is fixed to the same ratio as described by \cite{ref_uchiyama_2013}, and the overall normalization is set as a free parameter.

\begin{table*}[] \centering
\begin{minipage}{16cm}
\scriptsize
\caption{\centering Best-fit Parameters for Background Spectra}
\begin{tabularx}{16cm}{CCC|CCC}
\hline
\multicolumn{3}{c}{{\bf Region A (FI chips)}} & \multicolumn{3}{c}{{\bf Region B (BI chips)}}\\
 {\bf Model function}&  {\bf Parameter}&{\bf Value}& {\bf Model function}&  {\bf Parameter}&{\bf Value}\\
 \hline
 APEC(FE$_{\rm low}$) & VEM\footnotemark[a]($\rm cm^{-3}$)&$9.28^{+0.53} _{-0.54}\times 10^{59}$&APEC(FE$_{\rm low}$) & VEM\footnotemark[a]($\rm cm^{-3}$)&$9.56\pm^{+2.03} _{-2.03}\times 10^{58}$\\
$\rm constant_{1}$& $\rm factor_{1}$& $\rm 2.72^{+0.05}_{-0.06}$&$\rm constant_{1}$&${\rm factor_{1}}$&$\rm 17.85^{+0.23}_{-0.23}$\\
exponential& factor & 3.81 (fix)&bkn2pow&PhoIndx1& 1.60(fix)\\
& norm & 0.15 (fix)&&BreakE1 (keV)& 0.50(fix)\\
power-law& $\Gamma$ & 8.59 $\times$ $10^{-2}$ (fix)&&PhoIndx2& 0.46(fix)\\
& norm & 2.92 $\times$ $10^{-2}$ (fix)&&BreakE2 (keV)& 4.58(fix)\\
$\rm Gaussian_{1}$& $E_{1}$ (keV)& $\rm 1.82^{+0.01}_{-0.01}$&&PhoIndx3& 1.51(fix)\\
& $\sigma_{1}$ (keV)&$\rm 7.47^{+1.45}_{-1.31}$ $\times$ $10^{-2}$&&norm&  1.43 $\times 10^{-3}$(fix)\\
& norm & $\rm 1.67^{+0.21}_{-0.23}$ $\times$ $10^{-2}$ &$\rm Gaussian_{broad}$& $E$ (keV)& $\rm 13.17^{+0.06}_{-0.25}$ \\ 
$\rm Gaussian_{2}$& $E_{2}$ (keV)&  $\rm 8.77^{+0.11}_{-0.08}$&& $\sigma$ (keV)& $\rm 2.66^{+0.08}_{-0.07}$\\
& $\sigma_{2}$ (keV)& $\rm 1.11^{+0.09}_{-0.10}$&& norm & $\rm 0.19^{+0.02}_{-0.01}$\\
& norm & $\rm 0.11^{+0.01}_{-0.01}$&$\rm Gaussian_{1}$& $E_{1}$ (keV)& $\rm 1.90^{+0.02}_{-0.01}$\\
$\rm Gaussian_{3}$& $E_{3}$ (keV)& $\rm 2.16^{+0.01}_{-0.01}$&& $\sigma_{1}$ (keV)& 0.0 (fix)\\
& $\sigma_{3}$ (keV)& $\rm 4.71^{+0.99}_{-0.96}$ $\times$ $10^{-2}$&& norm & $\rm 1.72^{+0.39}_{-0.49}$ $\times$ $10^{-4}$\\
& norm & $\rm 1.95^{+0.08}_{-0.15}$ $\times$ $10^{-2}$&$\rm Gaussian_{2}$& $E_{2}$ (keV)& $\rm 1.77^{+0.11}_{-0.11}$\\ 
$\rm Gaussian_{4}$& $E_{4}$ (keV)& $\rm 7.46^{+0.00}_{-0.00}$&& $\sigma_{2}$ (keV)& 0.0 (fix)\\
& $\sigma_{4}$ (keV)& 0.0 (fix)&& norm & $\rm 7.26^{+0.47}_{-0.41}$ $\times$ $10^{-4}$\\
& norm & $\rm 2.61^{+0.11}_{-0.11}$ $\times$ $10^{-2}$&$\rm Gaussian_{3}$& $E_{3}$ (keV)& $\rm 2.15^{+0.01}_{-0.01}$\\
$\rm Gaussian_{5}$& $E_{5}$ (keV)& $\rm 9.70^{+0.00}_{-0.00}$&& $\sigma_{3}$ (keV)& $\rm 6.14^{+1.14}_{-1.13} \times 10^{-2}$\\
& $\sigma_{5}$ (keV)& $\rm 5.28^{+0.70}_{-0.64} \times 10^{-2}$&& norm & $\rm 5.59^{+0.46}_{-0.53}$ $\times$ $10^{-4}$\\
& norm & $\rm 8.29^{+0.23}_{-0.25} \times 10^{-2}$&$\rm Gaussian_{4}$& $E_{4}$ (keV)& $\rm 7.46^{+0.01}_{-0.01}$\\
&&&& $\sigma_{4}$ (keV)& 0.0 (fix)\\
&&&& norm & $\rm 2.84^{+0.33}_{-0.34}$ $\times$ $10^{-4}$\\
&&&$\rm Gaussian_{5}$& $E_{5}$ (keV)& $\rm 9.71^{+0.01}_{-0.01}$\\
&&&& $\sigma_{5}$ (keV)& $\rm 3.75^{+1.45}_{-1.88} \times 10^{-2}$\\
&&&& norm & $\rm 9.00^{+0.95}_{-0.89}$ $\times$ $10^{-4}$\\
\hline
\label{tab_background_part_result}
\end{tabularx}
\footnotetext[a]{Volume emission measure (VEM) is defined as $\int n_{\rm e}N_{\rm H}~dV$, where $n_{e}$, $N_{\rm H}$, and V are the electron density, the hydrogen density, and the emitting volume, respectively.}
\end{minipage}
\end{table*}

\begin{table}[] \centering
\begin{minipage}{8cm}
\scriptsize
\caption{\centering Best-fit Parameters of W50-originated components}
\begin{tabular}{ccc}
\hline
 {\bf Model function}&  {\bf Parameter}& {\bf Value} \\
 \hline
\multicolumn{3}{c}{\bf Region A}\\
TBabs&$N_{\rm H}$ ($\rm 10^{22} cm^{-2}$)&$1.75^{+0.10}_{-0.09}$\\
pegpwrlw&$\Gamma$&$2.99\pm 0.05$\\
 &norm \footnotemark[a]& $10.46^{+0.27}_{-0.24}$\\
 NEI& $kT$ (keV)& $0.22^{+0.004}_{-0.02}$\\
 & O (solar)& $0.27^{+0.38}_{-0.26}$\\ 
  & Ne (solar)& $0.49^{+0.16}_{-0.12}$\\ 
   & Mg (solar)& $0.76^{+0.17}_{-0.16}$\\ 
    & other elements (solar)& 1 (fixed)\\ 
  & $\tau$ ($\rm 10^{10} s~cm^{-3}$ ) & $8.46^{+6.43}_{-2.32}$\\
    & VEM\footnotemark[b] ($\rm cm^{-3}$)& $9.41^{+4.70}_{-3.26}\times 10^{58}$\\
    \\
&$\chi^2$/d.o.f \footnotemark[c]&781.46/638\\
\hline
\multicolumn{3}{c}{\bf Region B}\\
TBabs&$N_{\rm H}$ ($\rm 10^{22} cm^{-2}$)&$1.33^{+0.19}_{-0.22}$\\
pegpwrlw&$\Gamma$&$1.42\pm 0.10$\\
 &norm \footnotemark[a]&$ 2.93\pm 0.16$\\
 NEI& $kT$ (keV)& $0.93^{+0.17}_{-0.13}$\\
  & O (solar)& 0.27 (fixed)\\ 
  & Ne (solar)& 0.49 (fixed)\\ 
   & Mg (solar)& 0.76 (fixed)\\ 
     & other elements (solar)& 1 (fixed)\\ 
       & $\tau$ ($\rm 10^{10} s~cm^{-3}$ ) & 8.46 (fixed)\\
    & VEM\footnotemark[b] ($\rm cm^{-3}$)& 1.89$^{+0.92}_{-0.64}\times 10^{56}$\\
    \\
&$\chi^2$/d.o.f \footnotemark[c]&659.69/642\\
\hline
\end{tabular}
\label{tab_entire_bestfit}
\footnotetext[a]{The normalizations correspond to the flux, which is integrated over the energy range of 1.0--7.0~keV in this analysis.}
\footnotetext[b]{Volume emission measure (VEM) is defined as $\int n_{\rm e}N_{\rm H}~dV$, where $n_{e}$, $N_{\rm H}$, and V are the electron density, the hydrogen density, and the emitting volume, respectively.}
\footnotetext[c]{d.o.f: degree of freedom.}
\end{minipage}
\end{table}

We found that the obtained spectra require extra components for both Regions~A and B.
Since excess emissions are seen around line energies of highly ionized elements of O, Ne, and Mg, we speculate that a thermal component is required in addition to the expected non-thermal emission.
We thus assume a source emission model consisting of power law and non-equilibrium ionization (NEI; $vnei$ model in XSPEC) components.
We adopt the Tuebingen-Boulder absorption model \citep[tbabs;][]{ref_wilms_2000} for the interstellar absorption.
Because of limited statistics, we only set the abundance for O, Ne, and Mg, which emit prominent lines, as free parameters and fixed the other elements to the solar values.
The photon index $\Gamma$, electron temperature $kT$, and ionization timescale $\tau$ are also set as free parameters.
Since the elemental abundance and $\tau$ for Region~B cannot be constrained due to large statistical errors, we fixed these parameters to those of Region~A.
The best-fit parameters of the background and source emission are shown in tables~\ref{tab_background_part_result} and \ref{tab_entire_bestfit}, respectively.
We find that the photon indices are substantially larger in Region~A ($\Gamma=2.99$) than in Region~B ($\Gamma=1.42$), suggesting a spatial variation of the non-thermal emission.

\begin{figure}
 \begin{center}
\includegraphics[width=75mm]{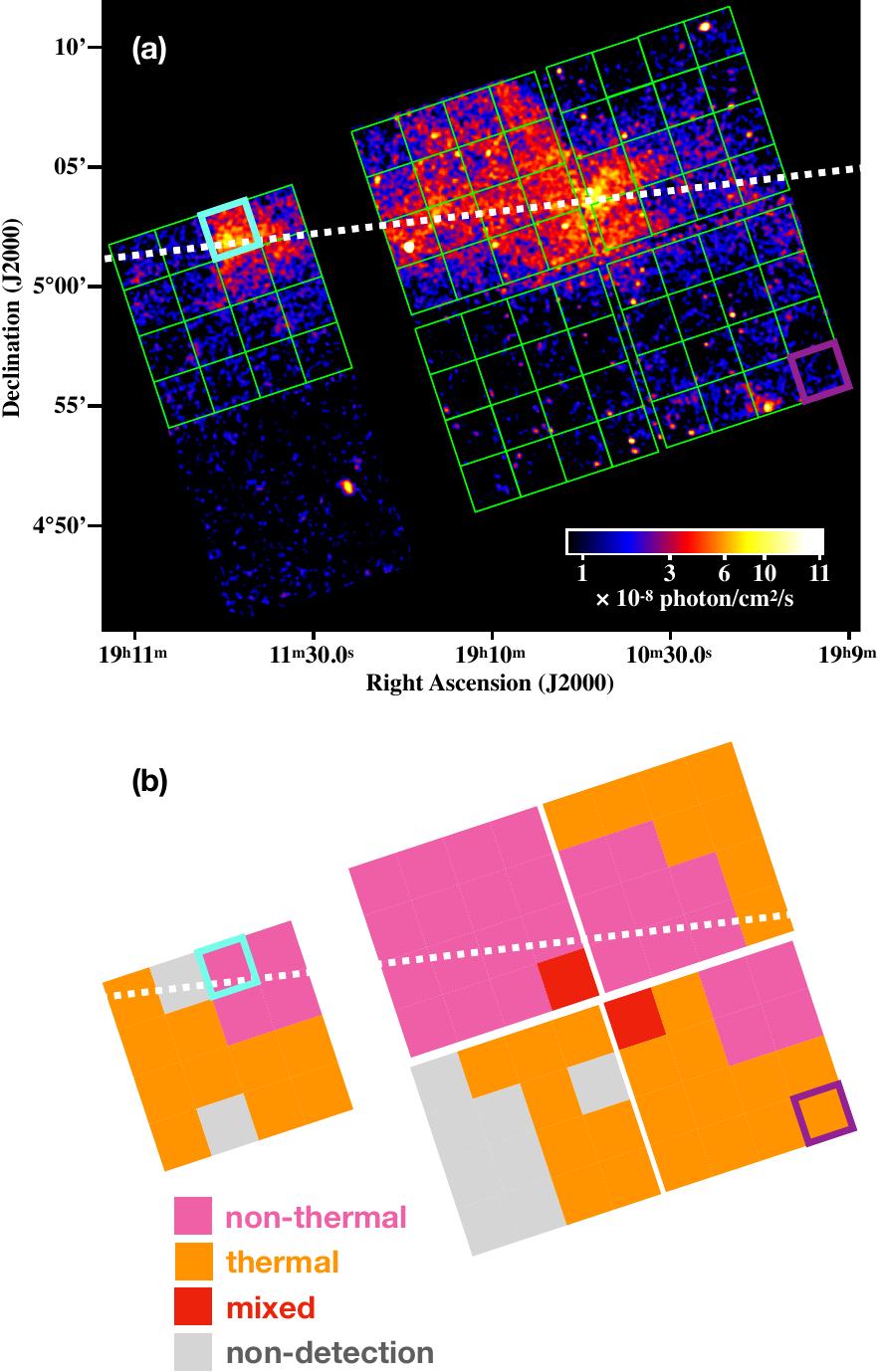} 
\end{center}
\caption{(a) Chandra ACIS image in the energy band of 0.5--7.0 keV. The green boxes indicate the spectra extraction regions for the analysis in \S\ref{sec_spatially}. The white dashed line denotes the jet precession axis. (b) Distribution of the thermal and non-thermal emissions.
The regions in magenta and orange correspond to those regions fitted well only with the non-thermal and thermal components, respectively.
The red regions require both components. No significant emissions from W50 are detected in the gray regions.}\label{fig_img_dist}
\end{figure} 

We evaluate the impact of systematic uncertainties in the sky background model on our analysis results.
The model by \citet{ref_uchiyama_2013} predicts that, inside the FOV of our data, the normalization of the GRXE changes by about $\pm5 \%$ as a function of Galactic longitude and latitude.
According to figure~4 of \cite{ref_uchiyama_2013}, furthermore, the normalization of GRXE is estimated to fluctuate by about $\rm \pm 20~\%$ around the model curves.
The normalization of CXB has uncertainties of $\rm \pm 14~\%$ as reported by \cite{ref_kushino_2002}.
We artificially change the normalizations of these components to the minimum and maximum within the range and analyze the spectra again.
We find that there are no significant differences in the analysis results, and conclude that those systematic uncertainties are negligible.
The fluctuation of the FE flux has not been studied well enough yet.
According to Table~4 of \cite{ref_ryu_2009}, the normalizations of the FE varies only about $\rm 20 \%$ in multiple regions within about $\rm \timeform{1.0D}\times\timeform{0.5D}$.
\cite{ref_uchiyama_2013} analyzed the spectra obtained from the Galactic center region and ridge region, and found that the intensities of the FE are almost constant in both regions \citep[see figure~7;][]{ref_uchiyama_2013}.
However, the fluctuation of the FE flux in the W50 region is uncertain.
Since the normalizations of the FE, as shown in Table~1, are roughly consistent with that reported by \citet{ref_uchiyama_2013}, we assume that the intensity of the FE is almost uniform and that the fluctuation is small enough within our FOV.

\subsubsection{Spatially Resolved Spectroscopy}\label{sec_spatially}
\label{sec_nonthermal}

We perform a spatially resolved spectroscopy in order to examine the spatial variation of the spectral parameters of the W50 emission suggested in \S\ref{sec_analysis_1}. 
We divided the FOV of each ACIS chip into 4$\times$4 as displayed in figure~\ref{fig_img_dist}.
\begin{figure}
 \begin{center}
\includegraphics[width=80mm]{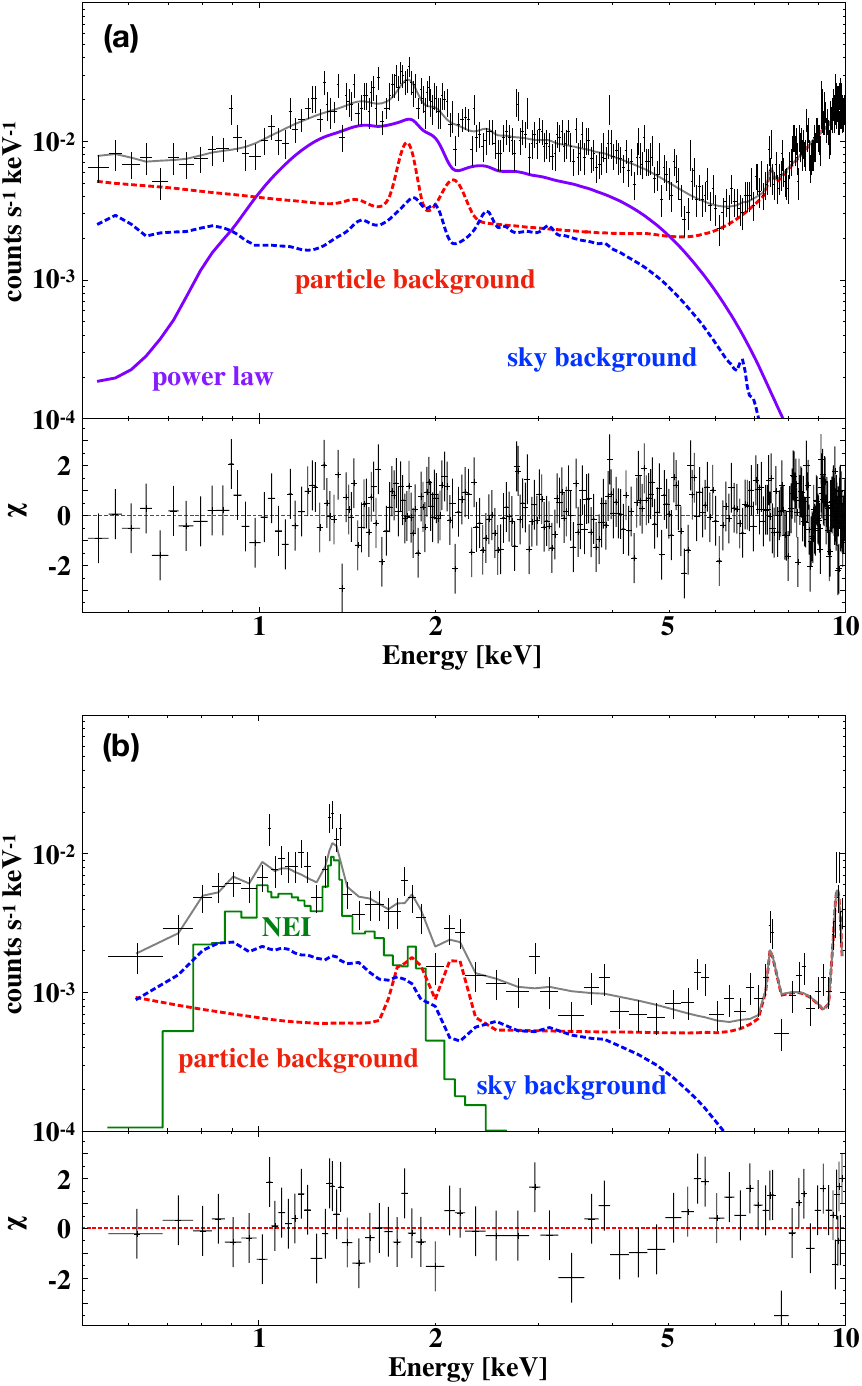} 
\end{center}
\caption{(a) Spectrum extracted from the region enclosed by the cyan lines in figure~\ref{fig_img_dist} as a non-thermal-dominant example. (b) Same but from the region enclosed by the purple lines in figure~\ref{fig_img_dist}, which is dominated by the thermal emission. The bottom panels present residuals from the best-fit models.
} \label{fig_img_dist_spec}
\end{figure} 

Extracted spectra are fitted with the same model as that used in \S\ref{sec_analysis_1}.
The elemental abundances and $\tau$ are fixed to the best-fit values in \S\ref{sec_analysis_1} (table~\ref{tab_entire_bestfit}) because of the limited data statistics.
We set $N_{\rm H}$, $\Gamma$, $kT$, and the normalizations as free parameters.
As described above, we assume the normalization of the FE is uniform over the FOV.
Therefore, the normalization of FE components is fixed at the value in Table~\ref{tab_background_part_result} scaled by area.
We found that spectra from some regions can be reproduced well only with either thermal or non-thermal components. 
With a criteria of $2\sigma$ significance for additional components, we categorized all the regions into four groups as shown in figure~\ref{fig_img_dist}.

\begin{figure*}
\begin{center}
\includegraphics[width=150mm]{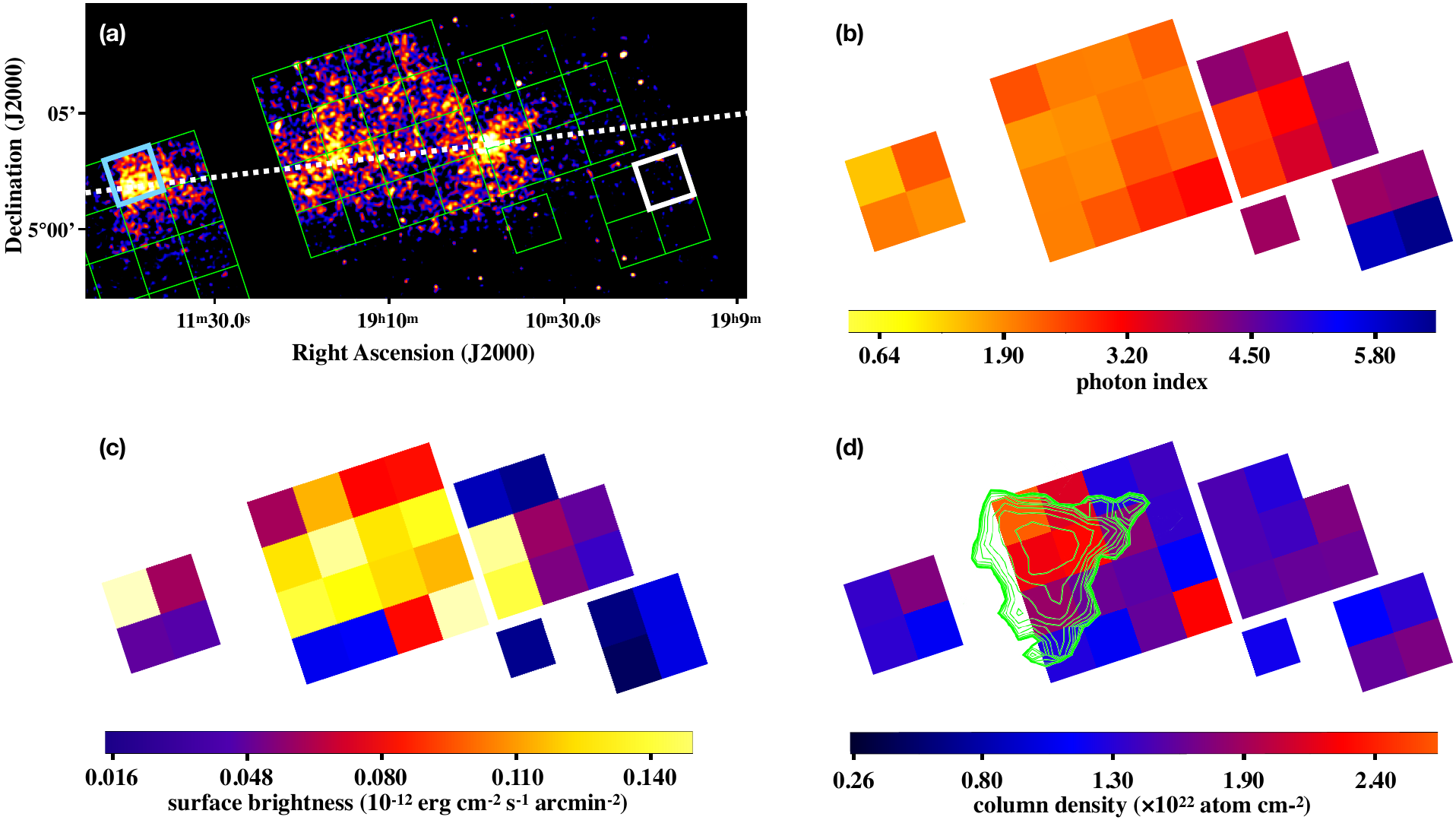} 
\end{center}
\caption{(a) Chandra ACIS image in the energy band of 2.0--7.0 keV.
(b) Photon index, (c) surface brightness integrated in 1.0--7.0 keV, and (d) column density maps. 
Overlaid on the map (d) are contours of $^{12}$CO emission integrated in the velocity range from 45~km~s$^{-1}$ to 60~km~s$^{-1}$ \citep{ref_yamamoto_ip}.}\label{fig_nonthermal_pow2}
\end{figure*}

The result in figure~\ref{fig_img_dist} clearly indicates that the non-thermal emission is dominant along the jet precession axis 
whereas the thermal emission is more uniformly distributed.
Example spectra are displayed in figure~\ref{fig_img_dist_spec}, in which we find clear spectral differences between the non-thermal and thermal dominant regions.
Figure~\ref{fig_nonthermal_pow2} shows parameter maps of the non-thermal emission according to the best-fit results. 
The non-thermal emission tends to have softer spectra as one moves away from SS 433 along the jet precession axis. 
This would be explained by synchrotron cooling as previous studies claimed \citep[e.g.,][]{ref_safi-harb_1997,ref_namiki_2000,ref_brinkmann_2007}.

\begin{figure}
 \begin{center}
\includegraphics[width=75mm]{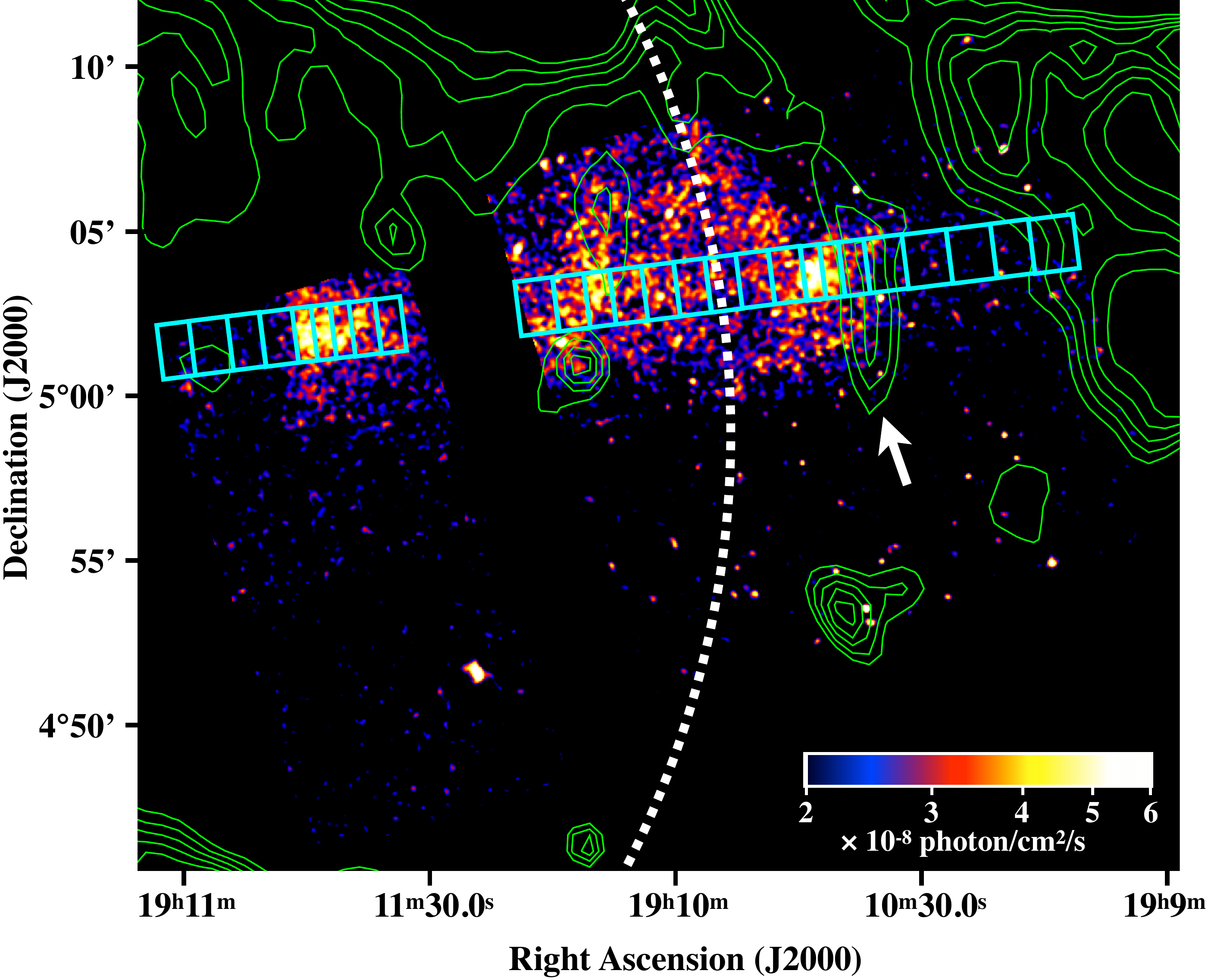} 
\end{center}
\caption{Chandra ACIS image in the energy band of 2.0--7.0 keV.
The cyan boxes indicate the spectra extraction regions for the analysis in \S\ref{sec_nonthermal}.
The green contours indicate the 140-MHz LOFAR radio continuum \citep{ref_broderick_2018}.
The white dashed line indicates the limb of the central shell of W50 as defined by \cite{ref_dubner_1998}.
The white arrow indicates "filament J" in \cite{ref_goodall_2011b}.
}\label{fig_img_l}
\end{figure}

\begin{figure}
\begin{center}
\includegraphics[width=75mm]{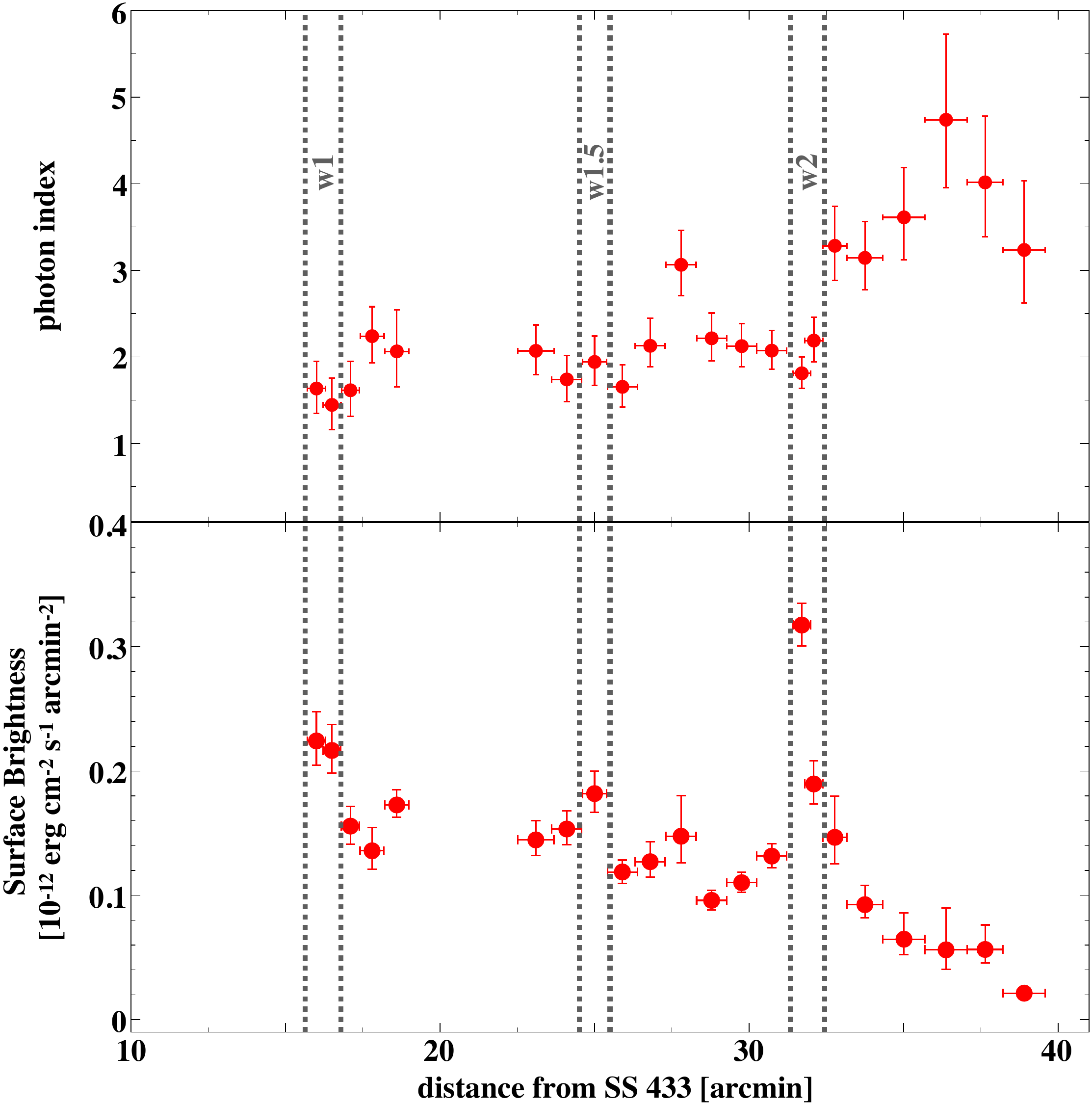} 
\end{center}
\caption{The photon index (top panel) and surface brightness in 1.0--7.0 keV (bottom panel) plotted against the distance from SS~433.
The locations between dashed lines correspond to the X-ray bright knots, w1, w1.5, and w2.}\label{fig_nonthermal_data}
\end{figure}

In order to obtain a more detailed view of the spectral variation along the jet precession axis, we perform additional spatially resolved analysis by extracting spectra from the regions shown in figure~\ref{fig_img_l}.
The X-ray radiations from the region on the jet precession axis are dominated by the non-thermal component as shown in figure~\ref{fig_img_dist}.
We apply the power-law model described above, and obtained spatial variations of the photon index and surface brightness as functions of distance from SS~433 as presented in figure~\ref{fig_nonthermal_data}.
Note that no significant non-thermal emission is detected in the regions closer to SS~433 than the w1 knot. 
We also evaluated the impact of the systematic uncertainties in the background model on the results as in section~\ref{sec_analysis_1}, and concluded that these are negligible.


\section{Discussion}
\label{sec_discussion}
\subsection{Thermal Emission}
As described in the previous section, the thermal emission seems more diffusely distributed than the non-thermal emission concentrating along the jet precession axis. 
While it is obvious that the non-thermal emission is directly related to the jet activity, the origin of the thermal emission is still uncertain.
The best-fit parameters of $kT$ and $\tau$ for Region~A (table~\ref{tab_entire_bestfit}) are roughly consistent with those obtained in the eastern lobe of SS~433 \citep[$kT \sim 0.3 $~keV and $\tau \sim \rm10^{12}$~s~$\rm cm^{-2}$;][]{ref_brinkmann_2007}, implying that W50 is filled with a thermal emission of a common origin.
However, we cannot immediately conclude that it is due to shock-heated interstellar material (ISM) since the abundances of O and Ne are significantly lower than the solar values, even taking into account the systematic uncertainties of the background model described in the section~\ref{sec_analysis_1}.
A possibility of metal-rich ejecta is also unlikely because the ionization timescale does not match the estimated density ($\sim$1~$\rm cm^{-3}$) and the age ($\sim$20000~yr) of W50 \citep{ref_goodall_2011}.

In the spatial distribution of the spectral components (figure~\ref{fig_img_dist}b), we found that the thermal emission is present also around the jet precession axis, particularly behind the knot w2, suggesting that the thermal emission is not only heated by the SNR shell but also related to the jet activity.
If this is the case, the jet-SNR interaction \citep{ref_zealey_1980,ref_downes_1986,ref_murata_1996} is a plausible scenario to account for the observational fact.
Hydrodynamical simulations \citep{ref_zavala_2008,ref_ohmura_2021} and radio/X-ray observations \citep{ref_dubner_1998,ref_brinkmann_2007} support the idea that the jet is penetrating the SNR shell of W50.
We, in principle, agree with this picture and consider that the jet penetrates the shell of ISM.
The elemental abundances variations of the ISM in the Galaxy were reported by \cite{ref_decia_2021}.
If the abundances of the ISM in the W50 region are locally lower than that of the solar values, the observation results may be explained with this picture.
We also propose an alternative scenario where the SNR shell is composed of swept-up circumstellar materials (CSM) produced by a stellar wind.
Low-abundant CSM shells are found in core-collapse SNRs such as RCW~103 \citep{ref_frank_2015} and 1E~0102.2$-$7219 \citep{ref_alan_2019}.
If a similar CSM shell was energized by the jet of SS~433, both low abundance and large spatial extension could naturally be explained.
More accurate measurement of the abundances in future observations \cite[such as XRISM;][]{ref_tashiro_2018} will nail down the origin of the shell and the progenitor mass. 

\subsection{Non-Thermal Emission} 
We have analyzed the spectra of the western lobe and found that the radiation is dominated by the non-thermal component as shown in figure~\ref{fig_img_dist}.
The best-fit $\Gamma$ for Region~B (table~2) is roughly consistent with that in the innermost part of the eastern lobe obtained by \cite{ref_safi-harb_1999}.
The radiation from the eastern lobe is interpreted as synchrotron radiation because of the photon index of 1.3--1.6, which is within the range expected for typical synchrotron radiation \citep{ref_safi-harb_1999}.
Therefore, the non-thermal emission in the western lobe can also be attributed to synchrotron radiation.
No non-thermal emission is detected from the region between SS~433 and the knot w1, and the synchrotron emission shows spectral steepening as away from SS~433, as shown in figure~\ref{fig_nonthermal_data}.
This can be explained if electrons are accelerated to relativistic energies at w1 and lose their energies through synchrotron cooling as they travel along the jet. 

\begin{figure*}
 \begin{center}
\includegraphics[width=170mm]{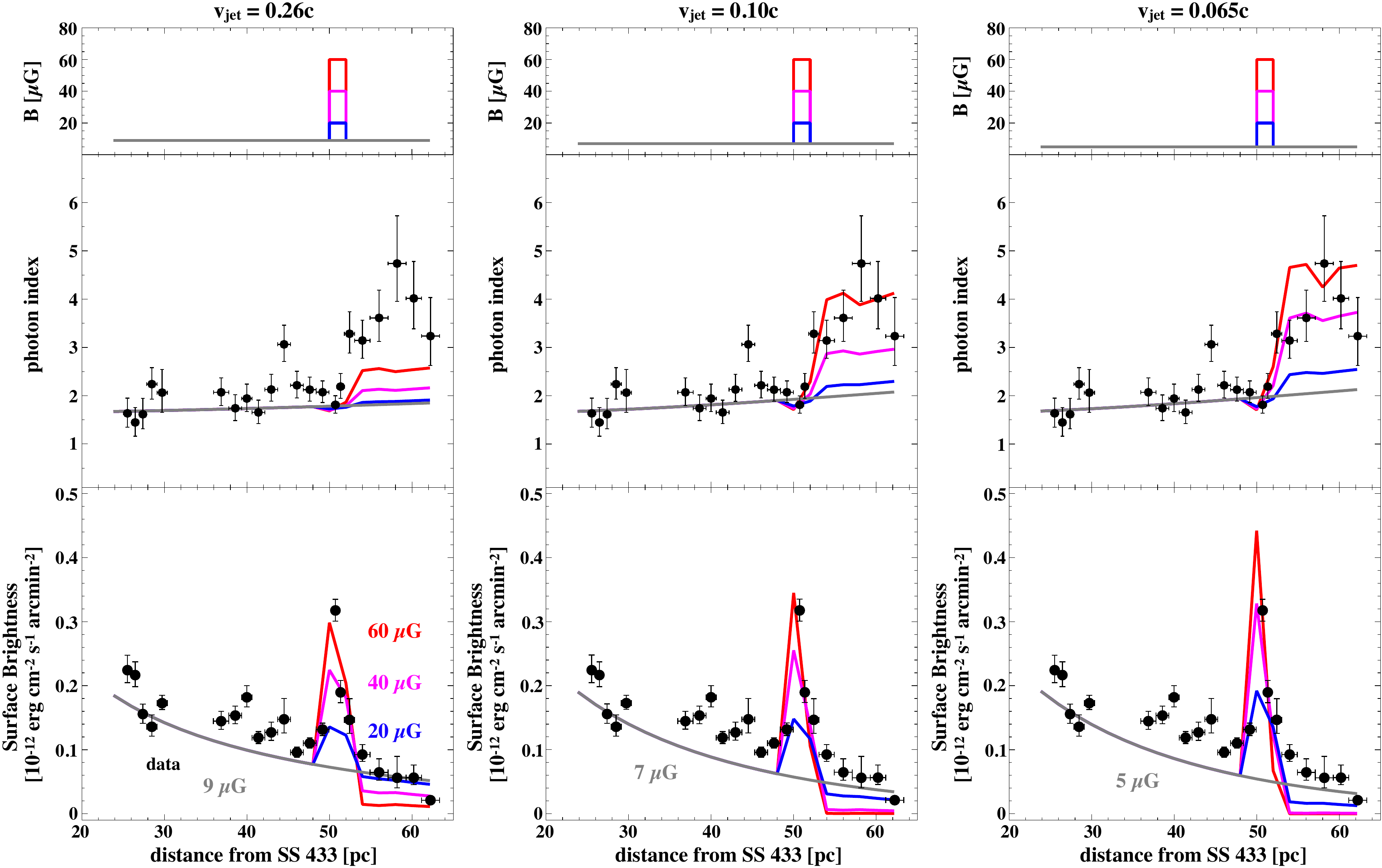} 
\end{center}
\caption{Comparison between the model and the observed spectral variation along the jet precession axis. The top panels indicate magnetic field strength profiles assumed in the model.
The middle and bottom panels present photon index and surface brightness profiles, respectively. 
The gray curves show models with uniform magnetic field strengths ($\rm 9~\mu G$, $\rm 7~\mu G$, and $\rm 5~\mu G$). 
The blue, magenta, and red curves indicate profiles at w2 with $\rm20~\mu G$, $\rm40~\mu G$, and $\rm60~\mu G$, respectively. 
The left, center, and right panels present model curves with $v_{\rm jet} = 0.26c$, $0.10c$, and $0.065c$, respectively.}\label{fig_nonthermal_pow}
\end{figure*}

Figure~\ref{fig_nonthermal_data} shows that the synchrotron spectra suddenly become steeper immediately outside the knot w2 like a step function.
Also, the surface brightness becomes higher at the location of w2. 
If particles are injected with a constant injection rate as a function of time into the acceleration site at the knot w1 and propagate through the jet \citep{ref_watson_1983,ref_margon_1984}, the observed structure of the knot w2 can be naturally explained with a locally strong magnetic field \citep{ref_sudoh_2020}.
The spectral slope variation can also be explained qualitatively since electrons are quickly cooled through synchrotron in the enhanced magnetic field.
The knot w1.5 would be another such region but without significant spectral slope changes, suggesting modest magnetic field strength there. 
Another possibility, we cannot rule out re-acceleration of particles at w2.
Since the photon index at w2 is $\Gamma = 2.18$, which can be translated into the slope of the injection spectrum $p_{\rm inj} = 3.36$, the particles may be re-accelerated by a low Mach number ($\rm M<2$) shock \citep[e.g.,][]{ref_gaisser_1990}.

\subsection{Model for Spectral Variation of Non-Thermal Emission}
We attempt to reproduce the spectral variation of synchrotron X-rays along the jet precession axis (figure~\ref{fig_nonthermal_data}) with a model developed based on the work by \cite{ref_sudoh_2020}. 
In the model calculation, we pay particular attention to the knot w2, for which our analysis result suggests magnetic field enhancement. 
The details of the model computation are described in Appendix. 
In the model, non-thermal electrons, which are responsible for synchrotron X-rays, are injected (i.e., accelerated) at the location of the knot w1. 
The injection spectrum of electrons is assumed to have a form of a power law with an exponential cutoff. 
The index of the injection spectrum is set to $p_{\rm inj} = 2.28$, corresponding to a photon index of synchrotron emission of $\Gamma = 1.64$, which is indeed obtained at the knot w1 in our spectral analysis. 
We assume the cutoff energy of the injection spectrum to be $E_{\rm cut} = 2~\mathrm{PeV}$ so that it is consistent with the result by \cite{ref_sudoh_2020}.
We note that this assumption is not sensitive for our purpose as long as $E_{\rm cut} \gtrsim 100~\mathrm{TeV}$. 
We adopt the same parameters of  the Galactic radiation field for inverse Compton (IC) scattering as those by \cite{ref_sudoh_2020}.

We try three different values for the jet velocity $v_{\rm jet}$: (i) the jet keeps its initial velocity ($v_{\rm jet} = 0.26c$), (ii) $v_{\rm jet} = 0.10c$ as estimated by \cite{ref_panferov_2017} for the bright knot in the eastern lobe at a distance from SS~433 similar to that of the western lobe, and (iii) $v_{\rm jet} = 0.26c/4 = 0.065c$, assuming a strong shock at knot w1.
Figure~\ref{fig_nonthermal_pow} shows the result of the model calculations plotted together with the observational data.
We find that, in the regions outside w2, the magnetic field strengths of $\rm 9~\mu G$, $\rm 7~\mu G$, and $\rm 5~\mu G$ well reproduce the data in the cases of $v_{\rm jet} = 0.26c$, $0.10c$, and $0.065c$, respectively. 

The magnetic field is required to be locally enhanced to fit the harder and brighter synchrotron emission at  the knot w2. 
In the case of $v_{\rm jet} = 0.26 c$, the model curve with a magnetic field strength of $\rm 60~\mu G$ nicely reproduces the surface brightness at the knot w2, but fails to reproduce the spectral slope, making $v_{\rm jet} = 0.26c$ an unlikely option. 
The model curves show excellent agreement with the observation both for photon indices and surface brightness are in excellent agreement with the observation with magnetic field strengths at w2 of $\rm 60~\mu G$ and 40~$\rm \mu G$ for $v_{\rm jet} = 0.10c$ and $0.065c$, respectively.
These results strongly indicate that $v_{\rm jet} \lesssim 0.10c$.
The magnetic field should be enhanced at the knot; a faster jet requires a stronger magnetic field at the knot.

\subsection{Origin of Knots}
\begin{figure}
 \begin{center}
\includegraphics[width=85mm]{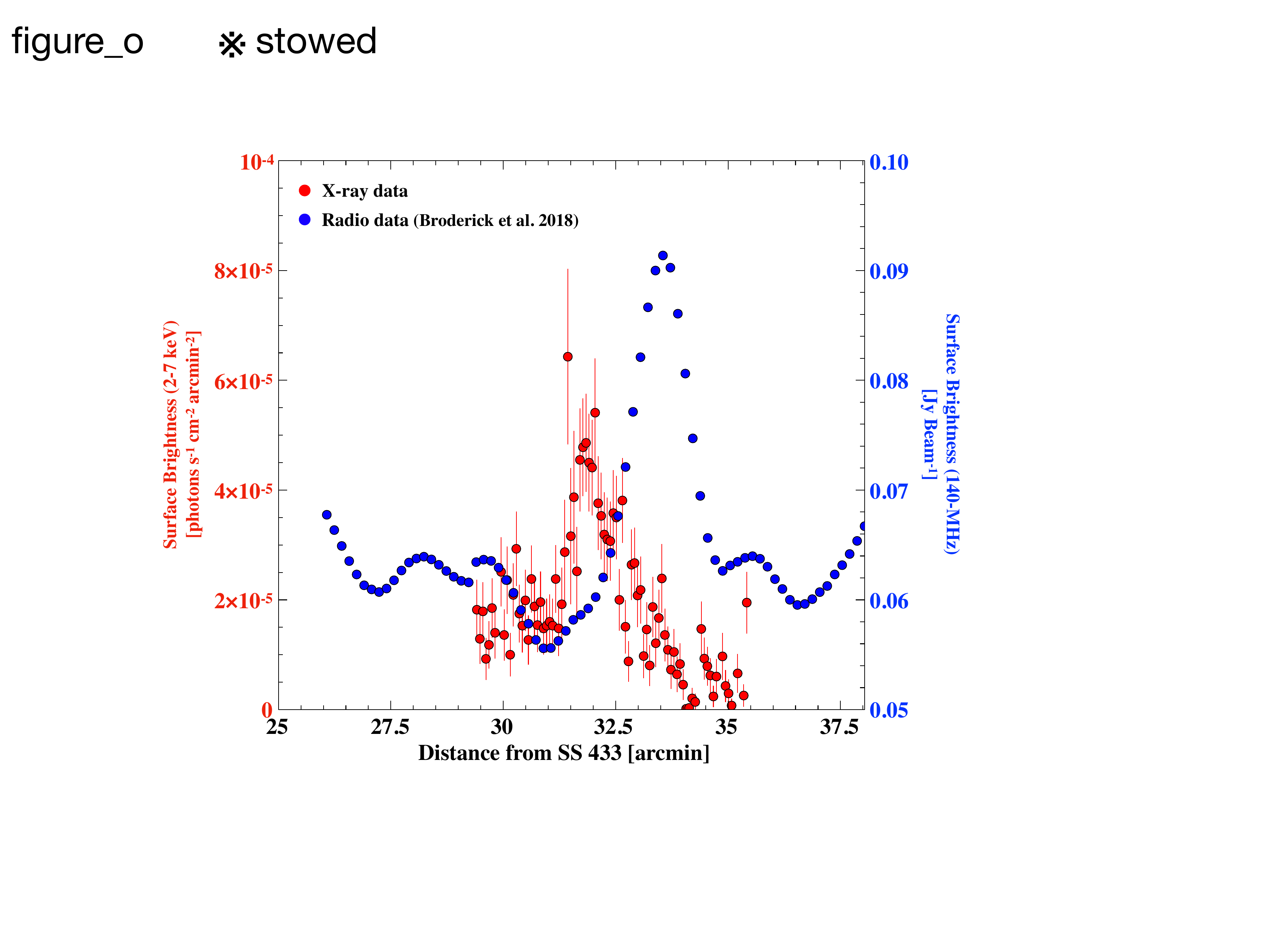} 
\end{center}
\caption{Surface brightness profiles of X-rays (red) and radio (blue) along the jet precession axis. The beam size is 78~arcsec~$\times$~55 arcsec.
}\label{fig_knotposition}
\end{figure}

\begin{figure}
 \begin{center}
\includegraphics[width=85mm]{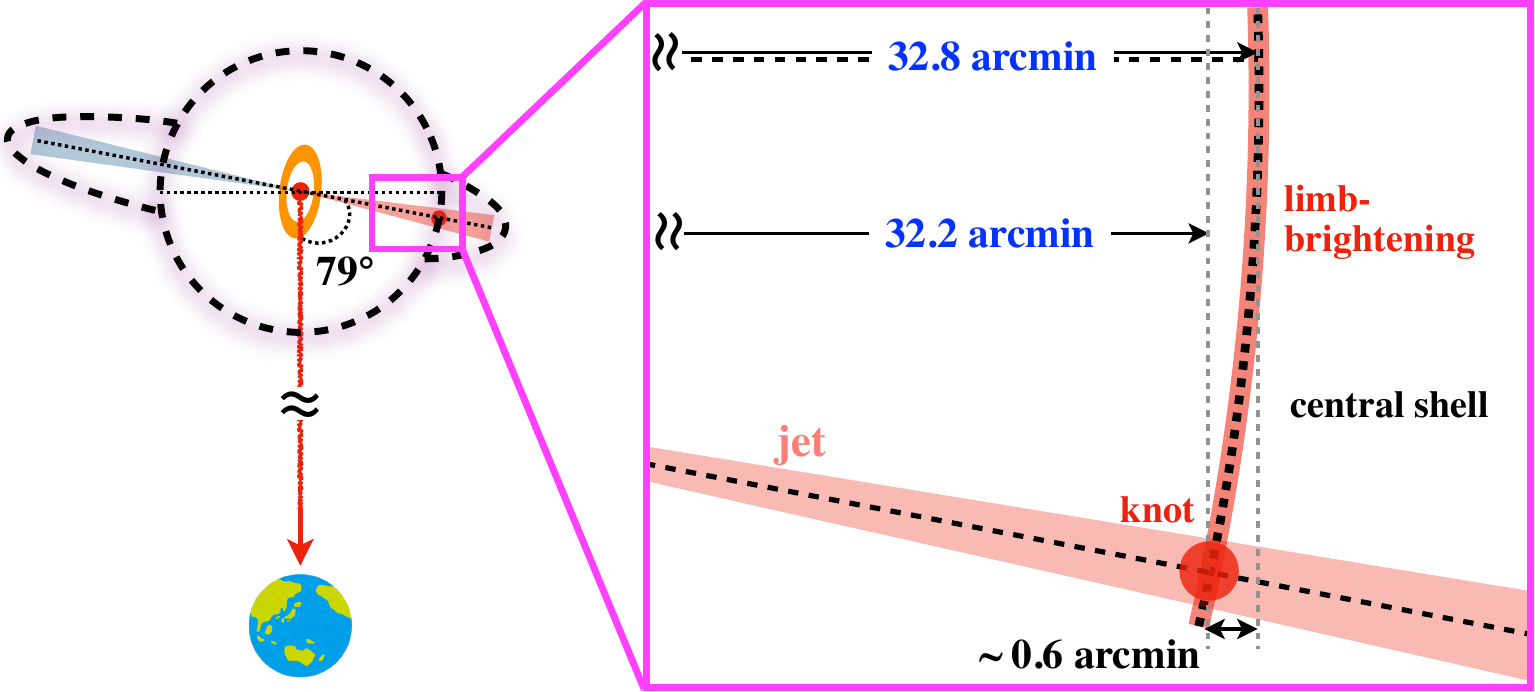} 
\end{center}
\caption{Proposed geometry of the SS~433/W50 system. 
The distance is assumed to be 5.5~kpc.
The inclination of the jet to the line of sight is assumed to be 79$^\circ$.}\label{fig_schematic} 
\end{figure}

As we discussed above, magnetic field enhancement is one possible scenario for explaining the harder and brighter synchrotron emission at the knot w2. 
Then, some mechanism to enhance the magnetic field must be at work.  
The comparison between X-ray and radio images (figure~\ref{fig_img_l}) provides some insights into this question. 
A filament-like radio structure is located close to the X-ray knot w2 \citep[``filament J" in][]{ref_goodall_2011b}. 
The surface brightness profiles plotted in figure~\ref{fig_knotposition} show that the radio filament is $\sim \timeform{1.50'}$ away from the knot.  
The radio structure is connected to the northern part of the central shell \citep[e.g., ][also see figure~\ref{fig_img_l}]{ref_dubner_1998,ref_broderick_2018}, and the two structures have similar surface brightness to each other.
These facts as well as the overall radio morphology prompt us to suggest that the two structures are different parts of the brightened limb of the same shell, i.e., the central shell. 
Under this hypothesis, we propose a geometry of the system as schematically drawn in figure~\ref{fig_schematic}. 
We refer to \cite{ref_margon_1989} and \cite{ref_eikenberry_2001} for the inclination of the jet to the line of sight, $79^{\circ}$. 
This may be different from the angle at the time when the jet reached the shell because of the jet precession with an angle of $\sim 20^\circ$ \citep{ref_margon_1989}. 
We assume the central shell is a perfect sphere, although, in reality, it may be distorted by the long-lasting jet--shell interaction.
Considering these uncertainties, the observed offset, $\sim \timeform{1.50'}$, is roughly consistent with the geometry in figure~\ref{fig_schematic}, which predicts the offset to be \timeform{0.6'}. 

If the jet is indeed interacting with the shell, one possible scenario is that the magnetic field at w2 is enhanced even before the jet penetrates the shell.
It is possible that the shell of W50 has strong magnetic fields of order $ \rm 10-100~\mu G$, like other middle-aged SNRs (e.g., W44: \citealp{ref_cardillo_2014}; W51c: \citealp{ref_koo_2010}).
The jet-shell interaction may have caused mixing of the jet and shell plasma and enhanced the magnetic fields at w2, assuming that the fields did not decay during this process.
Another possibility is that the jet-shell interaction amplified the magnetic field through compression.
Some instabilities may be triggered by the interaction, which can also amplify the magnetic field.

We have detected a new knot, w1.5, located between the previously known knots, w1 and w2 (figure~\ref{fig_img_b}). 
As shown in figure 6d, the location of the knot w1.5 coincides with the peak of the $^{12}$CO emission, which is from the molecular cloud, N4 \citep[][]{ref_yamamoto_2008,ref_su_2018,ref_yamamoto_ip}.
The figure also shows that $^{12}$CO emission has a clear positive correlation with X-ray absorption column densities. 
\cite{ref_yamamoto_ip} reported that the temperature of this cloud is higher than that of typical clouds in the Galactic plane, and a velocity shift of up to $\rm 3~km~s^{-1}$ from the velocity of the entire cloud ($\rm \sim 49~km~s^{-1}$) was detected only in the northwestern part. 
These results suggest that most of the cloud is located in front of the jet, and a part of the cloud is interacting with the jet.
Based on this evidence, we propose a possibility that the enhanced magnetic field at the location of the knot w1.5 is related to the interaction between the jet and the molecular cloud there.
We, however, note that as AGN jets propagate through the intercluster medium, the jets push away the surrounding medium and generate a hot cocoon around the jet \citep[e.g.,][]{ref_bromberg_2011}.
It is possible that the SS 433 jet also has a cocoon \citep[e.g.,][]{ref_asahina_2014}, and the molecular cloud is located within it.
The cloud must survive without fully evaporating under such an environment \citep[e.g.,][]{ref_cowie_1977,ref_liu_2020}.
In fact, \cite{ref_yamamoto_ip} argued from the intensity gradient of the $^{12}$CO emission that the envelope of the cloud has been stripped away and that only the dense part of the cloud survives.
It is beyond the scope of this paper to discuss further and will be studied in future works.

Unlike w2 and w1.5 discussed above, the knot w1 does not spatially coincide neither with a molecular cloud \citep{ref_yamamoto_2008} nor with the central shell of W50 \citep{ref_dubner_1998,ref_broderick_2018}.
The X-ray morphology suggests that the knot w1 corresponds to the location of electron acceleration. 
A question here is why particles are accelerated right at this location. 
A possible scenario is that particles are accelerated at internal shocks generated at w1 \citep[e.g.,][]{ref_rees_1978,ref_rees_1994,ref_kaiser_2000}.
However, the energy dissipation at the internal shock is inefficient due to the low velocity ($<0.26c$) of the jet \citep{ref_kobayashi_1997}.
Therefore, the internal shock model is unlikely.
Also, the re-collimation shock scenario is improbable for the precessing jets as described by \citet{ref_monceau-baroux_2014}.

We thus propose a scenario in which knot w1 is originated by the interaction between the jet and SNR shell.
\cite{ref_murata_1996} constructed a one-dimensional model that describes the interaction between the two, and showed that it generates forward and reverse shocks.
Based on their model, the locations of the shocks depend on the density of the jet and ISM.
For a jet-to-ISM density ratio of $\sim 1/9$, the observed location of the knot w1 is roughly consistent with the location of the reverse shock obtained from the simulation.
These results suggest that particles in the jet may be accelerated to very high energies by the reverse shock or by the shock wave generated by the interaction between the jet flow and the material swept up by the reverse shock.
Alternatively, the jet activity itself may generate a strong shock at the knot w1.
\cite{ref_ohmura_2021} simulated a scenario in which the SS~433/W50 system is formed by SS~433 jets themselves.
They found that particles are accelerated by the oblique shock generated by the interaction between the jet flow and backflows.
The synchrotron radiation intensity map of their simulation is roughly consistent with the observed X-ray structure (see figure~8 of \citealp{ref_ohmura_2021}).
Interestingly, the knot w1 and its counterpart in the east, e1, are located at almost the same distance from SS~433 \citep{ref_brinkmann_1996,ref_safi-harb_1997}, and so particle acceleration at both knots would be explained with a common scenario.

\section{Conclusions}
We performed a spatially resolved analysis of X-rays from the W50 western lobe with Chandra.
We found thermal emission distributing more diffusely than synchrotron X-rays, which are relatively concentrated along the jet precession axis. 
The parameters of the thermal emission are comparable to those in the eastern lobe reported by \cite{ref_brinkmann_2007}, 
implying that W50 is filled with a thermal emission of a common origin.
No significant synchrotron X-ray emission was detected from the region between SS~433 and the knot w1, suggesting electrons are accelerated there. 
The location of the knot w1 is roughly consistent with the simulation result of the location of the reverse shock generated by the jet-SNR interaction \citep{ref_murata_1996}.
Synchrotron X-ray spectra steepen as away from the knot w1, which can be explained by synchrotron cooling of electrons during their transport along the jet.
We also found that synchrotron X-rays show a rapid spectral steepening immediately outside the knot w2 ($\Gamma \sim 1.6$ to $\Gamma \sim 3$).
Comparison with the observational data and our model indicates that the knot corresponds to a region where the magnetic field is enhanced and electrons are quickly cooled.
The knot w2 spatially coincides with the radio shell of W50, suggesting the jet plasma is mixing with the shell, which has an enhanced magnetic field.
Alternatively, we propose scenarios in which the magnetic field is amplified by some mechanism (e.g., plasma compression; instabilities) near the central shell of W50.
Another new finding is the detection of the X-ray knot w1.5, which is at the location of the molecular cloud \citep{ref_yamamoto_ip}.
A possible origin of the knot is an enhanced magnetic field by a jet-cloud interaction.

\section{Acknowledgements}
We thank the anonymous referee for helpful comments to improve this manuscript.
We also thank Dr. J. W. Broderick for providing the LOFAR 140-MHz radio image.
This research has made use of data obtained from the Chandra Data Archive and the Chandra Source Catalog, and software provided by the Chandra X-ray Center (CXC) in the application packages CIAO, ChIPS, and Sherpa.
This work is supported by Grant-in-Aid for Japan Society for the Promotion of Science (JSPS) Fellows Number JP20J23327 (K.K.). 
This work is supported also by JSPS Scientific Research Grant Nos. JP19H01936 (T.T.), JP21H04493 (T.T. and T.G.T.), JP19K03915 (H.U.), JP18H05458 (Y.I.) and JP19K14772 (Y.I.).
Y.I. is also supported by World Premier International Research Center Initiative (WPI), MEXT, Japan.
T.S. is supported by JSPS Overseas Research Fellowships.

{
\appendix
\section*{One-Dimensional Model of Non-Thermal Particle Evolution}
\label{apen_model}
In section 3, we computed a model that predicts synchrotron spectra at each location of the SS~433 jet. 
The model is developed based on the work by \citet{ref_sudoh_2020}. 
We also refer to \citet{ref_hess_2020} for the treatment of radiative and adiabatic cooling effects on electrons. 
The energy--spatial density distribution of electrons at the injection site is assumed to be a power law with an exponential cutoff, 
\begin{equation}
n_0 = AE^{-p_{\rm inj}}\exp \left(-\frac{E}{E_{\rm cut}}\right),
\label{eq_1}
\end{equation}
where $E$, $E_{\rm cut}$, and $A$ are the electron energy, cutoff energy, and normalization, respectively. 
The parameter $z$ is the coordinate on the jet precession axis, corresponding to the distance from SS~433.
In this equation as well as in what follows, the subscript 0 denotes values at the location of the injection. 

After electrons are injected, they are transported along the jet with their distribution deformed by cooling.  
The energy--spatial density distribution of electrons at $z$ can be expressed as
\begin{equation}
n(z,E) = n_0(E=E_0)\frac{dE_0}{dE}\frac{dV_0}{dV} = n_0\varphi \frac{dE_0}{dE}, 
\label{eq_2}
\end{equation}
where $dV$ is the differential volume element of the jet.
The parameter $\varphi$ is the plasma compression, $\varphi=\rho/\rho_{\rm 0}$, where $\rho$ is the plasma density. 
When we assume a conical jet with a constant linear density as a function of $z$, $\varphi$ is inversely proportional to $z^2$.

The evolution of the electron energy is described by 
\begin{equation}
\frac{dE}{dz} = \frac{1}{v_{\rm jet}}\dot{E}(z, E). 
\label{eq_3}
\end{equation}
The energy loss rate $\dot{E}$ is the sum of radiative (synchrotron and inverse Compton) 
and adiabatic cooling rates: 
\begin{equation}
\dot{E}(z, E) = \dot{E}_{\rm syn}(z, E) + \dot{E}_{\rm IC}(z, E) + \dot{E}_{\rm ad}(z, E). 
\label{eq_4}
\end{equation}
The adiabatic cooling rate is written as
\begin{equation}
\dot{E}_{\rm ad}=\frac{v}{3} \frac{d\ln \rho}{dz}E.
\label{eq_5}
\end{equation}
The radiative cooling rate is 
\begin{equation}
\dot{E}_{\rm syn}(z, E) + \dot{E}_{\rm IC}(z, E) =-aE^2,
\label{eq_6}
\end{equation}
where $a$ is defined as
\begin{equation}
a = \frac{4}{3} \frac{\sigma_{\rm T}c}{(m_{\rm e}c^2)^2} (w_{\rm ph} + w_{\rm B}),
\label{eq_6_1}
\end{equation}
where $\sigma_{\rm T}$, $m_{\rm e}$, $w_{\rm ph}$, and $w_{\rm B}$ are the Thomson cross section, electron mass, seed photon energy density, and magnetic field energy density, respectively.
Using the magnetic field strength ($B$), $w_{\rm B}$ can be rewritten as $w_{\rm B} = B^2/8\pi$.
Then, equation (\ref{eq_3}) can be written as
\begin{equation}
v\frac{d}{dz}\left(\frac{\rho^{1/3}}{E} \right) = \rho^{1/3}(z)a(z). 
\label{eq_7}
\end{equation}
Solving this equation, we obtain 
\begin{equation}
n(z, E) = \varphi^{4/3}\left( \frac{E_0}{E} \right) n_0,
\label{eq_8}
\end{equation}
where $E_0$ is given as 
\begin{equation}
E_0 = E\frac{\varphi^{1/3}}{1-E\lambda \rho^{-1/3}}.
\label{eq_9}
\end{equation}
Here $\lambda$ is defined as
\begin{equation}
\lambda = \int_{z_{\rm s}}^{z} \rho^{1/3}(z')a(z')\frac{dz'}{v(z')}.
\label{eq_10}
\end{equation}
We then calculate synchrotron spectra with the emissivity formula described in \cite{ref_Ghisellini_1988}:
\begin{eqnarray}
\lefteqn{ \varepsilon(\nu,\gamma)=\frac{4\pi\sqrt{3}e^{2}\nu_{\rm B}}{c}\chi^{2}} \nonumber \\
& \ \ \ \ \ \ \ \ \ \ \times \left\{K_{\frac{4}{3}}(\chi)K_{\frac{1}{3}}(\chi)-\frac{3}{5}\chi \left[ K_{\frac{4}{3}}^{2}(\chi)-K_{\frac{1}{3}}^{2}(\chi) \right] \right\},
\label{eq_11}
\end{eqnarray}
where $\nu$, $\gamma$, and $K_{n}$ are the radiation frequency, the electron Lorentz factor, and the modified Bessel function of order $n$, respectively.
The parameters $\nu_{\rm B}$ and $\chi$ are defined as $\nu_{\rm B} \equiv eB/2\pi m_{\rm e}c$ and $\chi \equiv \nu/(3\gamma^{2} \nu_{\rm B})$, respectively.
}

\end{document}